\documentclass[12pt,tightenlines,a4paper]{article}
\usepackage{amssymb}
\usepackage{amsmath}
\usepackage{epsf}
\usepackage{axodraw}

\def\gappeq{\mathrel{\rlap {\raise.5ex\hbox{$>$}}
{\lower.5ex\hbox{$\sim$}}}}
\def\lappeq{\mathrel{\rlap{\raise.5ex\hbox{$<$}}
{\lower.5ex\hbox{$\sim$}}}}

\newcommand{\beq}{\begin{equation}}
\newcommand{\eeq}{\end{equation}}

\begin{document}
\oddsidemargin -0.4cm
\evensidemargin -0.8cm
\pagestyle{empty}

\begin{center}

\centerline{\Large\bf A stable hierarchy from Casimir forces }

\vspace*{5mm}
\centerline{\Large\bf and the holographic
interpretation } \vspace{2.0cm}

{\large Jaume Garriga$^{1,2}$ and Alex Pomarol$^{2}$}\\

\vspace{.8cm}
$^{1}$
Dept.  F{\'\i}sica Fonamental,
Universitat de Barcelona, Diagonal 647, 08028 Barcelona, Spain.

$^{2}$
IFAE, Universitat Aut{\`o}noma de Barcelona,
08193 Bellaterra (Barcelona), Spain.
\vspace{.4cm}
\end{center}

\vspace{1cm}
\begin{abstract}
We show that in the Randall-Sundrum model the Casimir energy
of bulk gauge fields
(or any of their supersymmetric relatives)
has contributions that depend logarithmically on the radion.
These contributions satisfactorily stabilize
the radion, generating a large hierarchy of scales
without fine-tuning.
The logarithmic behaviour
can be understood, in a 4D holographic description,
as the running of gauge couplings with the infrared
cut-off scale.
\end{abstract}
\vfill
\eject
\pagestyle{empty}
\setcounter{page}{1}
\setcounter{footnote}{0}
\pagestyle{plain}


\section{Introduction}

A very interesting approach to the hierarchy problem, which has
inspired a great deal of research in the past few years, is the
Randall-Sundrum (RS) model \cite{rs}. This consists of a 3-brane
of positive tension (Planck brane)
separated from a negative tension 3-brane (TeV brane) by a
slice of 5D Anti-de-Sitter (AdS$_5$) space. All mass scales in the theory are
comparable to some Planckian cut-off scale $M_P$, but energy
scales on the negative tension brane are redshifted (relative to
those on the positive tension brane) by a geometric ``warp
factor'' $ a=e^{-k \pi R}.$ Here $k \lesssim M_P$ is the inverse
of the AdS curvature radius and $R$ is the thickness of the
slice. By taking $R \approx 11 k^{-1}$, the mass scales on the
negative tension brane will be perceived as being only of the
order $aM_P\sim$ TeV, even if intrinsically they are comparable
to the cut-off scale. A central role in this model is played by
the interbrane distance $R$, since its value
 determines the
hierarchy.
The corresponding light degree of freedom (canonically
normalized) is known as the ``radion'' field $\phi$ \cite{gw}
\begin{equation}
\phi = f a \label{radion}=fe^{-k\pi R}\, .
\end{equation}
Here $f=\sqrt{6} M_P$, where $M_P\simeq 2.4\times 10^{18}$ GeV.
Clearly, a complete solution to the hierarchy problem in the context of the
RS model requires a stabilization mechanism for the radion.
Such mechanism should explain the extremely small value of the warp factor
$a\sim$ TeV$/M_P\sim 10^{-15}$
without the need of  introducing any small parameters.
Furthermore the mechanism should be stable under radiative corrections.

Soon after the RS model was proposed, Goldberger and Wise
\cite{gw} showed that the radion could be successfully stabilized
by introducing a classical interaction between the branes. For
this purpose, they considered a bulk scalar field $\Phi$ with mass
somewhat below the cut-off scale, and with potential terms on both
branes such that the vacuum expectation value
of $\Phi$ would be different on both
ends of the slice. The mechanism does not involve any fine-tuning or
exceedingly small parameters, and it gives the radion a mass
somewhat below the TeV.

Instead of introducing an ad-hoc classical interaction between the
branes, one may ask whether the
Casimir energy of bulk fields may be sufficient to stabilize the radion.
This issue was considered
in Ref.~\cite{gpt}, both for bulk conformal fields and for the graviton
field. In both cases, however, it was found that the Casimir
force would not stabilize the radion to the right value
unless a tuning of parameters was invoked. Even in this case,  the induced
radion  mass was too small.

Fortunately, this conclusion is not general. In this paper, we will show
that there are certain bulk fields whose contribution to the Casimir energy
density $V(a)$ depends logarithmically on $a$. In particular, gauge
fields (and their supersymmetric relatives)
induce terms of the form $V(a)\sim a^4/\ln a$,
which can stabilize $a$ at very small values. This is reminiscent of the
Coleman-Weinberg potential, where a small vacuum expectation value
can be obtained without the need of introducing any large or small parameters.
The interesting form of $V(a)$ can be understood by using the AdS/CFT
correspondence.
The RS model with gauge bosons in the bulk corresponds to
a 4D conformal field theory (CFT)
where a certain global symmetry has been gauged.
We will see that the logarithmic contribution to $V(a)$
arises from the running of the gauge coupling.

The plan of the paper is the following. In Section~\ref{sec2}, the
bulk fields giving logarithmic contributions to $V(a)$
will be identified by using the AdS/CFT
correspondence. In
Section~\ref{sec3}, the radion effective potential for these specific
bulk fields is computed in the slice of AdS$_5$.
The last Section is for conclusions.

\section{The radion potential from CFT}
\label{sec2}

Here
we will consider the 4D holographic
 version of the RS model (a slice of AdS$_5$) to
argue that there are sizable logarithmic contributions
to the radion potential.

The AdS/CFT correspondence relates a 5D theory
in AdS to a 4D strongly coupled CFT
 \cite{maldacena} with large number of ``colors'' $N$.
This is based on the identification of the
 5D fields at the AdS boundary,   $\Phi(x)$,
as the
sources of the CFT operators
\begin{equation}
 {\cal L}=\Phi(x) {\cal O}(x)\, .
\label{inter}
\end{equation}
The correspondence says \cite{adscft} that, for bosons, the
 masses $M$ of the 5D fields are related to the
dimension of the CFT operators according to
\begin{equation}
{\rm Dim}[{\cal O}]=\alpha+2\, , \ \ \ \ {\rm with}\ \ \ \ \
\alpha=\sqrt{\left(\frac{s}{2}\right)^2+
\frac{M^2}{k^2}}\, ,
\label{dicb}
\end{equation}
where $s=(4,2,4)$ for scalars, spin-1 fields and the graviton
respectively.
For fermions \cite{adscftf}, the  5D fields $\Psi$
at the AdS boundary must be
 subject to a projection  $\Psi_{\pm}=\frac{1}{2}(1\pm\gamma_5)\Psi$
for  $M\geq 0$ and $M<0$ respectively.
Only half of the degrees of freedom of the fermions are identified as
sources, ${\cal L}=\bar\Psi_{\pm}(x) {\cal O}_{\pm}(x)+h.c.$,
where
\begin{equation}
{\rm Dim}[{\cal O}_{\pm}]=\alpha_{\pm}+\frac{3}{2}\, , \ \ \ \
{\rm with}\ \ \  \ \
\alpha_{\pm}=\left|\frac{M}{k}\pm \frac{1}{2}\right|\, ,
\label{dicf}
\end{equation}
for $M\geq 0$ and $M<0$ respectively.

This correspondence can also be extended
to the case of a slice of AdS$_5$.
The Planck brane
corresponds to an ultraviolet  cutoff at $p=k\sim M_P$  in the 4D CFT \cite{gu},
 while the TeV brane
corresponds
to an infrared cutoff at $p=ak\sim$ TeV \cite{apr,rzp}.
Therefore a slice of AdS$_5$  corresponds to a slice
 of CFT  in the momentum-scale.
The ultraviolet cutoff   has two important  consequences.
First, higher dimensional terms, suppressed by the cutoff $k$,
can appear  in the theory.
Second,
the CFT  can
generate, via Eq.~(\ref{inter}),  kinetic terms for the sources $\Phi(x)$,
 which  become dynamical.
For the case where ${\cal O}$ is a conserved current,
this effect corresponds to a gauging of the associated conserved
symmetry in the CFT.
For example,  gravity and gauge theories in AdS$_5$ are dual to
a CFT coupled to
gravity and to gauge bosons in 4D \cite{gu,apr}.

Since the conformal symmetry is spontaneously broken in the infrared
\cite{apr,rzp},
a mass gap proportional to $ka$ is generated.
Bound states of CFT are expected to form,
analogous to the mesons and baryons in QCD.
Neglecting the
interaction with the sources,
the bound state masses scale linearly with $a$:
\begin{equation}
m_{\rm CFT}\simeq ka\, .
\label{massbs}
\end{equation}
However, once the interaction Eq.~(\ref{inter}) is taken into account,
the CFT bound states mix with the sources $\Phi$. This is analogous to
the mixing of the $\rho$ meson with the photon in QCD. The new
mass eigenvalues, which will now  have a nontrivial dependence on $a$,
are identified with the Kaluza-Klein (KK) masses in the slice of
AdS$_5$ \cite{apr}. In particular,
we shall be interested in the case where the correction to the mass behaves
logarithmically.

Let us consider first
the case of an abelian gauge theory in AdS$_5$.
In the CFT holographic picture, this corresponds to adding the perturbation
 ${\cal L}=g A_\mu(x) J^\mu(x)$, where Dim[$J$]=3.
The CFT gives a contribution to $\langle JJ\rangle$ dictated
by the conformal symmetry
 \begin{equation}
\langle JJ\rangle=c N p^2\ln p\, ,
\label{jj}
 \end{equation}
where $c$  is a $p$-independent constant.
The CFT correlator Eq.~(\ref{jj}) renormalizes the photon propagator.
Like in QED, this renormalization  can be incorporated
in the running of the gauge coupling:
 \begin{equation}
g^2(\mu)=\frac{g^2(k)}{1+cNg^2(k)\ln\frac{k}{\mu}}
\sim \frac{1}{c N\ln\frac{k}{\mu}}\, ,
\label{running}
\end{equation}
where in the last approximation we have taken the large $N$ limit.
As mentioned above, the photon mixes with the spin-1 CFT bound-states $V$.
Assuming that $g$ is small, we can treat this mixing as a perturbation,
and estimate  the ${\cal O}(g^2)$
correction  to the mass of $V$ (given by Eq.~(\ref{massbs})).
This correction arises from  the diagram of Fig.~\ref{massV}, which  gives
 \begin{equation}
\delta m^2_V\sim g^2(\mu\sim m_V) f^2_Vm^{2}_V\, ,
\end{equation}
where $f_V\propto \langle 0| J|V\rangle$.
In a large $N$, we have \cite{largen}  $f_V\sim\sqrt{N}$ and
we obtain for small $a$
 \begin{equation}
\delta m^2_V\sim  \frac{k^2 a^2}{\ln a}\, .
\end{equation}
We see that the  mass of $V$ has acquired
 a logarithmic dependence on the warp factor
$a$. Therefore its contribution to the vacuum energy will be of the order
 \begin{equation}
V(a)\sim m_V^4\ln\frac{m_V}{\mu}\sim k^4a^4\left(1+\frac{c_1}{\ln{a}}\right)\, ,
\end{equation}
where $m^2_V\simeq m^{2}_{\rm CFT}+\delta m^2_V$, and
we have used $\mu\sim m_V$.
Going back to the 5D dual theory, we  conclude
that a gauge theory in a slice  of AdS$_5$ must generate
at the one-loop level a radion potential with terms of order $a^4/\ln{a}$.

\begin{figure}[t]
    \begin{minipage}[t]{0.5\linewidth}
        \begin{center}
        \begin{picture}(40,40)(0,0)
            \Photon(0,20)(40,20){3}{7}
            \Line(40,22)(80,22)
            \Line(40,18)(80,18)
           \Line(0,22)(-40,22)
            \Line(0,18)(-40,18)
            \Vertex(40,20){2}
            \Vertex(0,20){2}
            \Text(-45,22)[r]{$V$}
            \Text(84,22)[l]{$V$}
            \Text(50,27)[br]{$gf_V$}
            \Text(10,27)[br]{$gf_V$}
        \end{picture}
        \caption{Correction to the mass of $V$.}
\label{massV}
        \end{center}
    \end{minipage}
    \begin{minipage}[t]{0.5\linewidth}
        \begin{center}
       \begin{picture}(40,40)(0,0)
            \DashLine(0,20)(40,20){3}
             \Line(18,18)(22,22)
             \Line(18,22)(22,18)
            \Line(40,22)(80,22)
            \Line(40,18)(80,18)
           \Line(0,22)(-40,22)
            \Line(0,18)(-40,18)
            \Vertex(40,20){2}
            \Vertex(0,20){2}
            \Text(-45,22)[r]{$S$}
            \Text(84,22)[l]{$S$}
            \Text(55,27)[br]{$\lambda k f_S$}
            \Text(10,27)[br]{$\lambda k f_S$}
        \end{picture}
        \caption{Correction to the mass of $S$.}
\label{massS}
\end{center}
    \end{minipage}
\end{figure}

Similar arguments can be applied also to massless scalars $\phi$
coupled to CFT operators,
 ${\cal L}=\lambda \phi(x) {\cal O}(x)$.
If  Dim[${\cal O}$]=3,  then
$\lambda$ is dimensionless and
is renormalized by the CFT in the same way as the gauge coupling $g$.
We  have then the same situation as the one of the photon
described above and we  expect that  terms of  order $a^4/\ln{a}$
will be induced in the radion potential.
In the 5D dual picture, we find from
Eq.~(\ref{dicb}) that  a scalar operator of Dim[${\cal O}$]=3
is  associated
with   a bulk scalar with  mass $M^2=-3k^2$.
The massless scalar $\phi$
must arise from
the 5D scalar field after dimensional reduction.
In order to have a massless mode in the spectrum,
the scalar field must have
a boundary  mass term equal to \cite{gp}
$[2k^2\delta(y)-2k^2\delta(y-\pi R)]\phi^2$.
We must notice, however, that this massless scalar is not stable
under radiative corrections, unless supersymmetry is invoked.
This problem does not arise for the fermions.
We can have   a massless chiral fermion $\Psi_+$ coupled
to  a CFT operator
${\cal L}=\lambda \bar \Psi_{+}(x) {\cal O}_+(x)$
with  Dim[${\cal O}_+$]=5/2.
Again
$\lambda$ is dimensionless and logarithmically dependent on the momentum,
guaranteeing that the vacuum energy has terms proportional to $a^4/\ln{a}$.
From Eq.~(\ref{dicf}),  this corresponds to a  bulk fermion
with  mass $M=k/2$ in AdS$_5$.

It should be noted that the three cases
presented above
are related by supersymmetry \cite{gp}.
In AdS$_5$ the photon is in a  vector supermultiplet that contains
 a fermion of  mass $M=k/2$.
On the other hand, a fermion of
 mass $M=k/2$ can also be contained in a hypermultiplet
with
 a scalar of bulk mass $M^2=-3k^2$ and  boundary  mass term
$[2k^2\delta(y)-2k^2\delta(y-\pi R)]\phi^2$.
This scalar,  fermion and the photon
correspond to the three cases found above.
All of them have $\alpha=1$ [see Eq.~(\ref{dicb})].
In  the vector multiplet,   there is also a scalar
of mass $M^2=-4k^2$ and Dirichlet boundary conditions
(which therefore has no 4D massless mode).
This  has $\alpha=0$.
By supersymmetry this case must also induce  a $V(a)$
with terms $\sim a^4/\ln a$.
To analyze this last case, let us go back to the CFT holographic picture.

Let us consider a  scalar, $\phi$, with mass of order $k$,
 coupled to  a CFT operator of
Dim[${\cal O}$]=2:
 ${\cal L}=\lambda k \phi(x) {\cal O}(x)$.
Since  the coupling is of order $k$, an unsuppressed
correction to the mass of a scalar CFT  bound-state $S$
 can be generated
from
the  diagram of Fig.~\ref{massS}.
The correction is of order
 $\delta m^2_S\sim\lambda^2 f_S^2m_S^2$ where $f_S\propto\langle 0|{\cal O}|
S\rangle$.
The conformal symmetry  guarantees that $\lambda$ evolves logarithmically
with the momentum.
Therefore
$\delta m^2_S$ and one-loop vacuum energy
induced by this state have a logarithmic dependence on $a$.
Contrary to the above cases, in which we needed dynamical sources
to generate the diagram of Fig.~\ref{massV}, here
the scalar $\phi$ can be considered an auxiliary (nondynamical) field
\footnote{The kinetic term of $\phi$ does not play any role
in the correction to the mass of $S$.
The effect of $\phi$ is equivalent to inserting   a
dimension $4$ operator ${\cal L}=\lambda^2(\mu){\cal OO}$
at the cutoff scale.}.
In the  AdS picture, this effect will come from a
bulk scalar of mass $M^2=-4k^2$
with boundary conditions such that the dimensionally reduced 4D theory
 has no massless modes, e.g.,
Dirichlet boundary conditions.

\section{Radion effective potential in RS}
\label{sec3}

As mentioned in the introduction, the radion  develops an
effective potential at the one-loop level due to the Casimir energy of bulk
fields. This effect has been studied in Ref.~\cite{gpt}, for the case
of bulk conformal fields and  the graviton. More
general bulk fields were considered in Refs.~\cite{casimir}. However,
it appears that the interesting behavior of the specific bulk
fields mentioned in the previous section has been overlooked so
far.

The starting point for the calculation of the effective potential
is the spectrum of KK states of the different fields, as a
function of the radion expectation value $a$.
This spectrum depends on
the spin, as well as on the boundary conditions on the branes. In
general, it is given by the roots of the following
equation \cite{rs,gp}
\begin{equation}
F(m_n)=  J^{T}_{\alpha}(m_n/a k) Y^{P}_{\alpha}(m_n/k)-
J^{P}_{\alpha}(m_n/ k) Y^{T}_{\alpha}(m_n/a k) = 0\, .
\label{spectrum}
\end{equation}
The superindices $T$ and $P$ stand for TeV and Planck branes
respectively. For fields with Neumann boundary conditions, the
functions $J^i_\alpha$ are defined as
\begin{equation}
 J_{\alpha}^{i}(z) =  \epsilon_{i} J_\alpha(z) + z
J_{\alpha-1}(z), \quad \quad (i=T,P)\label{jtilde}
\end{equation}
where $J_{\alpha}$ are the usual Bessel functions and
\begin{eqnarray}
\epsilon_{i}\equiv& {s\over 2}-\alpha -r_{i}\, ,  &\ \ {\rm for\
bosons}\, ,\label{epsilono}\\ \epsilon_{i}\equiv&  {s\over
2}-\alpha_\pm \pm \frac{M}{k}\, ,
  &\ \ {\rm for}\ \Psi_{\pm}\ {\rm fermions}\, .
 \label{epsilon}
\end{eqnarray}
The parameters $\alpha$ and $s$ are given in Eqs.~(\ref{dicb}) and
(\ref{dicf}). The functions $ Y^{i}_{\alpha}$ are defined by a
similar equation, with $Y_{\alpha}$ replacing $J_{\alpha}$. The
parameters $r_T$ and $r_P$ represent the mass terms on the TeV and
on the Planck brane respectively.  For instance, the mass term for
a boson $\Phi$ including the brane contributions takes the form
$[M^2 + 2 k r_P \delta(y) - 2 k r_T \delta(y-\pi R)]\Phi^2$. In
the case of gravitons ($\alpha=2$) and massless vectors
($\alpha=1$) we have \cite{gpt,gp} $\epsilon_{i}=0$ and therefore
$ J_{\alpha}^{i}(z) = z J_{\alpha-1}(z)$. The same applies to the
corresponding fermionic superpartners \cite{gp}.

Eqs. (\ref{epsilono}-\ref{epsilon}) can be easily generalized to
include all sorts of local terms on the brane. For instance,
derivative terms are easily incorporated by replacing the
constants $\epsilon_i$ by functions of $z$. A massless gauge field
with brane kinetic terms would have
\begin{equation}
\epsilon_i = -b_i z^2, \label{turrons}
\end{equation}
where $b_i$ are constants. As we shall see, terms of this sort can
in some cases change the sign of the Casimir interaction.

The case with Dirichlet boundary conditions can be obtained from
the previous expressions by taking the limit
\begin{equation}
\epsilon_{P}\to - \infty\, ,\ \ \ \ \ \ \ \
\epsilon_{T}\to  \infty\, .
 \label{dirichlet}
\end{equation}
For bosons this corresponds to  infinite masses   on the brane.
In the limit Eq.~(\ref{dirichlet})
the second term in Eq.~(\ref{jtilde}) becomes negligible and we have $
J_{\alpha}^{i}(z) = J_{\alpha}(z)$, and $ Y_{\alpha}^{i}(z) =
Y_{\alpha}(z)$, where we have dropped an irrelevant multiplicative
constant.

From Eq.~(\ref{spectrum}) we can get the  masses for the KK states of the
photon. As we said, they correspond to the CFT bound-states
in the mass eigenstate basis.
In the limit $a\ll 1$ we obtain
\begin{equation}
m_n\simeq
\left(n-\frac{1}{4} \right)\pi ka-\frac{1}{\ln a }\frac{\pi ka}{2}\, ,\ \ \ \ \ \
n=1,2,...\, ,
\end{equation}
that shows the $a$-dependence  expected
from the CFT arguments given in Section~\ref{sec2}.

The sum over KK contributions to the effective potential for the
radion can be expressed as a contour integral in the complex $m_n$
plane, involving the logarithmic derivative of the function
$F(m_n)$. After some manipulations, this integral can be written
as \cite{gpt,casimir}
\begin{equation}
V(a) =  {k^4 \over 16\pi^2}\left[A + a ^4 B\right] + {\Delta
V}(a)\, , \label{veff}
\end{equation}
where
\begin{equation}
{\Delta V}(a) = (-1)^S g_* {k^4 a^4\over 16\pi^2} \int_0^{\infty}
dt\ t^3 \ln\left| 1 - { K^T_\alpha(t)  I^P_\alpha (a t) \over
I^T_\alpha(t) K^P_\alpha(a t)}\right|\, .
\label{calv}
\end{equation}
Here, $g_*$ is the number of physical polarizations, and $S=0,1$
for bosonic and fermionic degrees of freedom respectively. The
first two terms in Eq.~(\ref{veff}) correspond to
renormalizations of the Planck and TeV brane tensions,
respectively.
The piece proportional to $A$ is just a
constant term  which must be adjusted to give a zero
cosmological constant.
The functions in the integrand
in Eq.~(\ref{calv}) are given in terms of the Bessel functions
$I_\alpha$ and $K_\alpha$ through the following relations:
\begin{eqnarray}
 I^{i}_\alpha(\rho)& =& \epsilon_{i} I_\alpha(\rho) +
\rho I_{\alpha-1} (\rho)\, ,\\
 K^{i}_\alpha(\rho)& =& \epsilon_{i} K_\alpha(\rho) -
\rho K_{\alpha-1} (\rho)\, .
\end{eqnarray}

Let us now find the form of  $\Delta V(a)$  in the limit of a large
hierarchy $a\ll 1$.
The first thing to notice is that
$K^T_\alpha(t)/I^T_\alpha(t) \sim e^{-2t}$  for $t\gg 1$,
 and therefore, only the range
$t\sim 1$ will make a significant contribution to the
integral in Eq.~(\ref{calv}).
The dependence on the radion $a$ of this integral
arises from   ${ K^P_\alpha(a t)/I^P_\alpha(a t)}$,
which can be expanded for small $at$  as
\begin{equation}
{ K^P_\alpha(a t) \over  I^P_\alpha(a t)}
=\left( {
\Gamma(\alpha)\Gamma(1+\alpha)\over (\epsilon_P +
2\alpha)}\left({at\over
2}\right)^{-2\alpha}\left\{{\epsilon_P\over 2}  + \left({at\over
2}\right)^{2} {1\over 1-\alpha}\right\}- {\pi \over 2
\sin\alpha\pi}\right) \left[1+ O(a^2t^2)\right]\, .
 \label{expansion}
\end{equation}
Since $a$ is extremely small, and $\alpha>0$, the leading term is
proportional to $\epsilon_P$. Hence, the behavior will be very
different for $\epsilon_P=0$ or for $\epsilon_P\neq 0$.

Of particular interest will be the case of a massless gauge
boson ($\epsilon_P =0$, $\alpha=1$), and a scalar with
$M^2=-4k^2$ ($\epsilon_P \neq 0, \alpha=0$), where the right hand
side of Eq.~(\ref{expansion}) behaves
\footnote{This  can be seen from
the expansion  $a^{2(1-\alpha)}\approx 1+2(1-\alpha)\ln a$
and $a^{-2\alpha}\approx 1-2\alpha\ln a$ in the limit
$\alpha\rightarrow 1$ and $\alpha\rightarrow 0$
respectively.}
 as $\ln a$.
This will imply a
correction to the effective potential which behaves as $1/\ln a$,
as suggested by the AdS/CFT arguments presented in the previous
section. These two special cases will be dealt with in the
following subsections.

When $\alpha$ is not close to $0$ or $1$, one can show from
Eq.~(\ref{expansion}) that $\Delta V(a)$ behaves either as $a^4$ or as
a higher power of $a$. Terms proportional to $a^4$ can be absorbed
in a redefinition of $B$, whereas the higher powers have a
negligible effect. Hence, we shall disregard these cases in the
following discussion.

\subsection{Gauge fields ($\alpha=1$)}

As mentioned above, for a massless gauge field
we have $\epsilon_P =0$
and $\alpha=1$. In the limit $|\alpha-1|\ll |1/\ln a|$,
Eq.~(\ref{expansion}) reduces to
\begin{equation}
{ K^P_\alpha(a t) \over  I^P_\alpha(a t)}\approx \ln(a t/2) +
\gamma + O(a^2 t^2)\, ,
\end{equation}
where $\gamma$ is Euler's constant.
Since we are interested in the situation where $|\ln a|\gg 1$,
the argument of the logarithm in Eq.~(\ref{calv}) is close
to unity and we have
\begin{equation}
{V}(a) \approx {k^4 a^4\over 16\pi^2}\left(B -  {
g_*(-1)^{S}\beta_1(\epsilon_T) \over \ln a}\right) + const.,
\label{deltavap}
\end{equation}
where we have introduced
\begin{equation}
\beta_\alpha(\epsilon_T) \equiv \int_0^{\infty} dt\ t^3\ {
K^T_\alpha(t) \over  I^T_\alpha(t)}.\label{beta}
\end{equation}
The extremum of the potential is determined by the condition
\begin{equation}
\ln a \approx (-1)^{S} g_* {\beta_1(\epsilon_T)/ B}\,  ,
\label{minimum}
\end{equation}
that
corresponds to
a minimum if the
 radion mass
\begin{equation}
m^2_\phi \approx  {(-1)^{S} g_* \beta_1(\epsilon_T) \over (2 \pi \ln a)^2}\
{a^2 k^4 \over f^2}\, ,
\label{mass}
\end{equation}
is positive. Eq.~(\ref{minimum}) shows that a exponentially small
$a$ can be  obtained from parameters of order one. In particular
the large  $M_P/$TeV value
 can be obtained if the
r.h.s. of Eq.~(\ref{minimum})  takes the value  $\approx -35$. For
massless gauge bosons without any local terms on the brane
($\epsilon_T=0$) we have $\beta_1(0)\approx -1.005 < 0$ and the
radion mass, Eq.~(\ref{mass}), would be negative. This is easily
remedied by including the brane kinetic terms, in which case
$\epsilon_T$ has the form (\ref{turrons}). With a coefficient
$b_T\sim 1$, we have $\beta_1(b_T t^2)
> 0$ and of order unity. This can stabilize the radion, giving it a
mass $m_\phi\approx g_*^{1/2}a k^2/(2\pi f\ln a) \sim 4\times
10^{-3}g_*^{1/2}(ak^2/f)$. Other contributions which tend to
stabilize the radion come from the corresponding fermionic field
(with a mass $M=k/2$), which gives a positive $\beta_1$ when the
brane kinetic terms are small ($b_T\ll 1$). We may also consider
the situation with different boundary conditions.
If we keep the Neumann boundary condition on the Planck brane,
$\epsilon_P =O(a^2t^2)$, and impose Dirichlet conditions on the
TeV brane ($\epsilon_T \to \infty$), we also have
$\beta_1(\infty)\approx 2.33>0$. So, again, the boson tends to
stabilize the radion. In the 4D dual this last case would
correspond to breaking the gauge symmetry by giving to the gauge
boson a mass of order TeV.

\subsection{Fields with $\alpha\approx 0$}

From Eq.~(\ref{dicb}), a scalar field with $\alpha=0$ has a bulk
mass  $M^2=-4k^2$. If we add mass terms on the branes,
$r_P=r_T=2$ (so that $\epsilon_T=\epsilon_P=0$), then it is easy
to show from Eq.~(\ref{spectrum}) that there is a zero mode in the
spectrum, and all other mass eigenstates are positive. In general,
we will not have tachyons in the spectrum of KK masses provided
that
\begin{eqnarray}
&&\epsilon_T\epsilon_P \ln a + \epsilon_T- \epsilon_P \geq 0\, ,\label{notach1}\\
&&\epsilon_T \geq \epsilon_P\, .
 \label{notach2}
\end{eqnarray}
These two conditions are automatically satisfied if $\epsilon_T
\geq 0$ and $\epsilon_P \leq 0$,
which we now assume.
In the case where $\epsilon_P \not=0$ and $\alpha$ is small,
we have from
Eq.~(\ref{expansion}):
\begin{equation}
{ K^P_\alpha(a t) \over  I^P_\alpha(a t)}\approx
-\left[\ln(at/2)+\gamma+{(1/\epsilon_P)}+O(a^2t^2)\right]\, .
\end{equation}
Provided that $|\ln a+{1/\epsilon_P}| \gg 1$, the integral in
Eq.~(\ref{calv}) can be approximated as
\begin{equation}
{V}(a) \approx {k^4 a^4 \over 16\pi^2}\left(B + {g_*(-1)^{S}
\beta_0(\epsilon_T) \over \ln a+{1/\epsilon_P}}\right)+ const.,
\label{vapprox}
\end{equation}
where $\beta_0(\epsilon_T)$ is given in Eq.~(\ref{beta}).
The extremum of the potential is determined by the
condition
\begin{equation}
x\equiv \ln a+(1/ \epsilon_P)\approx (-1)^{S+1}g_*
{\beta_0(\epsilon_T)/B}, \label{xdef}
\end{equation}
and the radion mass at the extremum is given by
\begin{equation}
m^2_\phi \approx  {(-1)^{S+1} g_* \beta_0(\epsilon_T) \over (2 \pi
x)^2}\ {a^2 k^4 \over f^2}\, .
\label{radion2}
\end{equation}
A boson (S=0) with small $\epsilon_T$ can lead to a positive radion
mass since
$\beta_0(\epsilon_T\ll 1) \approx -\beta_1(\infty) \approx
-2.33$.
The mass will be largest for the smaller possible value of
$|x|$. Given that $\ln a \approx -35$, and since here we are
considering negative $\epsilon_P$, the best we can do is to
consider Dirichlet boundary conditions on the Planck brane
$\epsilon_P\to -\infty$, which corresponds to
$|x|\approx 35$.
The mass of the radion is then comparable to the one obtained in
Eq.~(\ref{mass}).

We may consider more general situations, where for instance
$\epsilon_P>0$, provided that we stay in the range of values of
$a$ where the conditions of Eqs.~(\ref{notach1}) and (\ref{notach2}) are
satisfied. In the general case, the approximation in Eq.~(\ref{vapprox})
will not be valid. However, it will still be true that $\Delta
V/a^4$ depends on $a$ through the combination $x$ defined in
Eq.~(\ref{xdef}). For $\epsilon_P>0$, the conditions
Eqs.~(\ref{notach1}) and (\ref{notach2}) require $x>0$, but in contrast with
the previous case, there is no requirement that $x$ should be
large. Hence, one may expect the existence of minima of $V(a)$ at
$x\sim 1$, where the mass of the radion is of order TeV without
the need of fine adjustments in the parameters. Since the analytic
approximation in Eq.~(\ref{vapprox}) breaks down, one has to proceed
numerically. For instance, we find that for $\epsilon_T=0.5$ and
$\epsilon_P \approx 1/39$, the potential has a minimum for
$B\approx 0.3 g_*$ where $x\sim 1$ and where the mass of the
radion is given by $m_\phi \approx 10^{-2}g_*^{1/2}(a k^2/f)$.
This is slightly larger than the mass in Eq.~(\ref{mass}), but not
dramatically so. Also, in this case one has to worry about the
neighboring regime of smaller values of $a$, where
Eq.~(\ref{notach1}) is violated. This could make
the minima with $x\sim 1$ metastable, which makes the case with
$\epsilon_P>0$ somewhat less appealing.

\section{Conclusions}

We have shown that in the RS model the radion can be stabilized by
the Casimir energy of certain bulk fields, leading to the observed
hierarchy $a\sim$TeV$/M_P$ without the need of introducing small
parameters. The main ingredient
 is the appearance of Casimir energy contributions that depend
logarithmically on the radion $a$. These contributions arise from
gauge fields in the bulk or from any of their supersymmetric
relatives: a fermion with mass $M=k/2$ or a scalar with mass
$M^2=-4k^2$. It should be emphasized that these bulk fields could
correspond to the 5D Standard Model gauge bosons or fermions.
Using AdS/CFT the logarithmic terms in $V(a)$ can be understood as
the running of the gauge coupling in the 4D dual theory.

Hence the Casimir force  provides a natural alternative to the
Goldberger-Wise mechanism for stabilizing the radion in the RS model,
as required for a complete solution to the hierarchy problem.

\section*{Acknowledgements}

We thank Santi Peris for useful discussions.
This work was partially supported by the CICYT
Research Projects
AEN99-0766, FPA2002-3598,
the MCyT and FEDER Research Project
FPA2002-00748, and DURSI Research Project 2001-SGR-00188.


\begin{thebibliography}{99}

\bibitem{rs}
L.~Randall and R.~Sundrum,
Phys.\ Rev.\ Lett.\ {\bf 83} (1999) 3370.

\bibitem{gw}
W.~D.~Goldberger and M.~B.~Wise,
Phys.\ Rev.\ Lett.\  {\bf 83} (1999) 4922;
Phys.\ Lett.\ B {\bf 475} (2000) 275.

\bibitem{gpt}
J.~Garriga, O.~Pujolas and T.~Tanaka,
Nucl.\ Phys.\ B {\bf 605} (2001) 192.

\bibitem{casimir}
W.~D.~Goldberger and I.~Z.~Rothstein,
Phys.\ Lett.\ B {\bf 491} (2000) 339;
D.~J.~Toms,
Phys.\ Lett.\ B {\bf 484} (2000) 149;
A.~Flachi and D.~J.~Toms,
Nucl.\ Phys.\ B {\bf 610} (2001) 144;
A.~Flachi, I.~G.~Moss and D.~J.~Toms,
Phys.\ Lett.\ B {\bf 518} (2001) 153;
A.~Flachi, I.~G.~Moss and D.~J.~Toms,
Phys.\ Rev.\ D {\bf 64} (2001) 105029;
I.Brevik, K. Milton,S. Nojiri and S. Odintsov, Nucl. Phys. B, {\bf
 599} (2001) 305;
S.~Nojiri, S.~D.~Odintsov and S.~Zerbini,
Class.\ Quant.\ Grav.\  {\bf 17} (2000) 4855;
J.~Garriga, O.~Pujolas and T.~Tanaka,
arXiv:hep-th/0111277;
W.~Naylor and M.~Sasaki,
Phys.\ Lett.\ B {\bf 542} (2002) 289;
I.~G.~Moss and W.~Naylor,
arXiv:hep-th/0106228;
A.~A.~Saharian and M.~R.~Setare,
arXiv:hep-th/0207138;
E.~Elizalde, S.~Nojiri, S.~D.~Odintsov and S.~Ogushi,
arXiv:hep-th/0209242.



\bibitem{maldacena}
J.~Maldacena,
Adv.\ Theor.\ Math.\ Phys.\ {\bf 2} (1998) 231.

\bibitem{adscft}
E.~Witten,
Adv.\ Theor.\ Math.\ Phys.\  {\bf 2} (1998) 253;
S.~S.~Gubser, I.~R.~Klebanov and A.~M.~Polyakov,
Phys.\ Lett.\ B {\bf 428} (1998) 105.

\bibitem{adscftf}
M.~Henningson and K.~Sfetsos,
Phys.\ Lett.\ B {\bf 431} (1998) 63;
W.~Muck and K.~S.~Viswanathan,
Phys.\ Rev.\ D {\bf 58} (1998) 106006.

\bibitem{gu}
See for example, S.~S.~Gubser,
Phys.\ Rev.\ D {\bf 63} (2001) 084017.


\bibitem{apr}
N.~Arkani-Hamed, M.~Porrati and L.~Randall,
JHEP {\bf 0108} (2001) 017.


\bibitem{rzp}
R.~Rattazzi and A.~Zaffaroni,
JHEP {\bf 0104} (2001) 021;
M.~Perez-Victoria,
JHEP {\bf 0105} (2001) 064.




\bibitem{gp}
T.~Gherghetta and A.~Pomarol,
Nucl.\ Phys.\ {\bf B586} (2000) 141.


\bibitem{largen}
E.~Witten,
Nucl.\ Phys.\ B {\bf 160} (1979) 57.



\end{thebibliography}
\end{document}